\documentclass[journal]{IEEEtran}
\usepackage{cite}
\ifCLASSINFOpdf
   \usepackage[pdftex]{graphicx}
\else
   \usepackage[dvips]{graphicx}
\fi
\usepackage[cmex10]{amsmath}

\ifCLASSOPTIONcompsoc
  \usepackage[caption=false,font=normalsize,labelfont=sf,textfont=sf]{subfig}
\else
  \usepackage[caption=false,font=footnotesize]{subfig}
\fi
\usepackage{algorithm}
\usepackage{algorithmic}
\usepackage{textcomp}
\usepackage{xcolor}
\usepackage{cite}
\usepackage{verbatim}

\ifCLASSINFOpdf
\else
\fi
\begin{document}
%
\title{A Trained Regularization Approach Based on Born Iterative Method for Electromagnetic Imaging}
%
%
%

\author{Abdulla Desmal,~\IEEEmembership{Member,~IEEE}
\thanks{A. Desmal is with the Department
of Electrical Engineering, Higher Colleges of Technology, RAK campus,
UAE e-mail: adesmal@hct.ac.ae}
\thanks{Manuscript received April XX, XXXX; revised September XX, XXXX.}}

%
%

\markboth{Journal of \LaTeX\ Class Files,~Vol.~XX, No.~X, September~XXXX}%
{Shell \MakeLowercase{\textit{et al.}}: Bare Demo of IEEEtran.cls for Journals}
%



\maketitle

\begin{abstract}
A trained-based Born iterative method (TBIM) is developed for electromagnetic imaging (EMI) applications. The proposed TBIM consists of a nested loop; the outer loop executes TBIM iteration steps, while the inner loop executes a trained iterative shrinkage thresholding algorithm (TISTA). The applied TISTA runs linear Landweber iterations implemented with a trained regularization network designed based on U-net architecture. A normalization process was imposed in TISTA that made TISTA training applicable within the proposed TBIM.  The iterative utilization of the regularization network in TISTA is a bottleneck that demands high memory allocation through the training process. Therefore TISTA within each TBIM step was trained separately. The TISTA regularization network in each TBIM step was initialized using the weights from the previous TBIM step. The above approach achieved high-quality image restoration after running few TBIM steps while maintained low memory allocation through the training process. The proposed framework can be extended to Newton or quasi-Newton schemes, where within each Newton iteration, a linear ill-posed problem is optimized that differs from one example to another. The numerical results illustrated in this work show the superiority of the proposed TBIM compared to the conventional sparse-based Born iterative method (SBIM).           
\end{abstract}

%
\IEEEpeerreviewmaketitle

\section{Introduction}
%
%
%
%
\IEEEPARstart{M}{achine} learning (ML) implementations in electromagnetic imaging (EMI) have gained broad interest through the past few years due to their efficiency \cite{xiao2020dual,yao2019two,mydur2001neural,li2018deepnis, wei2018deep, sandhu2019neural,li20213}. The developed schemes can be categorized into direct and indirect approaches. 

In the direct approaches, neural networks (NNs) are trained to retrieve the material properties directly from the measured scattered fields. In \cite{xiao2020dual}, a direct approach was proposed that utilized dual NNs. The first NN provided primary permittivity reconstruction using extreme learning machine, while the second NN was an enhancement convolutional neural network (CNN) that retrieved finner image details. In \cite{yao2019two}, another dual NNs approach was suggested, where the first model was a CNN followed by a deep residual CNN model. In \cite{mydur2001neural}, a multi-layer fully connected NN was designed that extracts the location and dimensions of an elliptical target. On the other hand, for indirect approaches, NNs are trained to extract inverse parameters or enhance existing reconstruction approaches based on first-order methods or recently based on fully nonlinear algorithms \cite{li2018deepnis, wei2018deep, sandhu2019neural,li20213 }. In \cite{li2018deepnis}, an indirect method known as DeepNIS was proposed that trained an image enhancement CNN using profiles reconstructed from a first-order backpropagation model. In \cite{wei2018deep}, a U-net filter was trained using a dataset collected from dominant current components that were extracted using the truncated singular value decomposition. In \cite{sandhu2019neural}, the number of nonzero pixels was predicted first using a fully connected NN and then utilized to select the kernel base of a modified sparse reconstruction greedy algorithm.  In \cite{li20213}, an expectation-maximization unsupervised learning approach was combined with the variational Born iterative model (VBIM) to classify the medium parameters of investigated targets.  

Iterative shrinkage thresholding algorithms (ISTAs) that use soft-thresholded Landweber iterations have been implemented successfully for sparse domain reconstruction \cite{daubechies2004iterative}. In \cite{desmal2014shrinkage}, ISTA was integrated into the Born iterative model (BIM) to develop a sparse BIM (SBIM). In \cite{desmal2014preconditioned}, ISTA was integrated into a preconditioned contrast-source inexact Newton framework. Recently, a trained ISTAs (TISTAs) have gained more interest among the image processing society due to their high quality in image restoration \cite{jin2017deep,ongie2020deep,gilton2019neumann}. The developed schemes use additive or projective regularization networks. The regularization networks were trained to learn the image features implicit in the training set. In TISTAs, the Landweber step length can be selected as a learning parameter. So far, the schemes that adopt TISTA consider a single linear operator independent of input images.                

In this work, a trained BIM (TBIM) is proposed that benefits from the recently developed TISTA. Since the first BIM iteration implements Born approximation for any applied example, TISTA experiences the same linear operator. Therefore, the TISTA framework mentioned above can be applied to reconstruct the first TBIM step. During the second TBIM step and after, the linear model is updated based on the retrieved contrast profile from the previous TBIM step. Hence, TISTA will experience different linear operators depending on the executed example. Consequently, using a single learning parameter for the step length yields divergence in the training process. Such argument occurs as well in Newton and quasi-Newton algorithms. In this work, the power iteration was implemented to normalize the linear operators that make TISTA training applicable. Since the trained regularization network is executed within each TISTA iteration step, the memory allocation during the training process might overflows, especially under large regularization networks, e.g., U-net networks \cite{ronneberger2015u}. In this work, TISTA within each TBIM step was trained separately, and the regularization network weights were initialized using the trained TISTA weights from the previous TBIM iteration step. The latter approach helped TBIM to converge in few steps while achieving high-quality image restoration in the testing set and low memory allocation throughout the training process.      

The remaining of the paper is organized as follows. Section~\ref{sec:CF} explains the contrast-field (CF) two-dimensional (2D) EMI forward formulation. Section~\ref{sec:sbim} outlines the SBIM, while Section~\ref{sec:tbim} introduces the proposed TBIM. Section~\ref{sec:results} illustrates the numerical results that compare the two schemes SBIM and TBIM, and test the performance of TBIM under different signal-to-noise ratios (SNRs).
\section{Formulation} \label{sec:formulation}
\subsection{Contrast-field equations} \label{sec:CF}
Assume a 2D investigated domain $S^{\rm inv}$ that lies in an infinite background free space medium, while $S^{\rm mea}$ is the measurement domain surrounding $S^{\rm inv}$, where $N^{\rm tra}$ line source transmitters and $N^{\rm rec}$ receivers are located. $S^{\rm inv}$ is discretized using $N$ equal square pixels of side length $\Delta d$. Let ${\rm \bf e}_{tr}^{\rm sca}$ be an $N^{\rm rec} \times 1$ vector that stores the scattered fields at the receiver locations due to transmitter $\{tr\}$. The data equation in the CF formulation computes ${\rm \bf e}_{tr}^{\rm sca}$ as follows      
\begin{equation} \label{eq:data}
    {\rm \bf e}_{tr}^{\rm sca}={\rm \bf G}{\mathcal D}({\rm \bf e}^{\rm tot}_{tr}) {\rm \bf t}
\end{equation}
where ${\rm \bf t}$ and ${\rm \bf e}^{\rm tot}_{tr}$ are $N\times 1$ vectors that store samples of the contrasts and total fields per transmitter $\{tr\}$, respectively, evaluated at the center of the discretized pixels. The operator $\mathcal D (.)$ transforms the input vector into a diagonal matrix. ${\rm \bf G}$ is an $N \times N^{\rm rec}$ matrix computed as follows  
\begin{equation} \label{eq:green_function}
{\rm \bf G}_{n,m}=k_{0}\int_{S_n} G({\rm \bf r}',{\rm \bf r}_m)d{s}'
\end{equation}
where $k_0$ is the background wavenumber, $S_n$ is the domain of the $n^{\rm th}$ pixel, and ${\rm \bf r}_m$ is a position vector that points toward the $m^{\rm th}$ receiver location. $G({\rm \bf r}',{\rm \bf r}_m)=H^{(2)}_{0}(k_{0}|{\rm \bf r}'-{\rm \bf r}_m|)/4j$ is the 2D scalar Green's function, where $H^{(2)}_{0}(.)$ is the second kind zero-order Hankel function. In the CF formulation, the state equation reads     
\begin{equation} \label{eq:stat}
    {\rm \bf e}^{\rm inc}_{tr}=\left ({\rm \bf I}+ {\rm \bf A} {\mathcal D}({\rm \bf t}) \right ) {\rm \bf e}^{\rm tot}_{tr}
\end{equation}
where ${\rm \bf e}^{\rm inc}_{tr}$ is an $N \times 1$ vector that stores incident field samples at the pixels' centers. ${\rm \bf A}$ is an $N \times N$ matrix computed as $\rm \bf G$ in (\ref{eq:green_function}) with ${\rm \bf r}_m \in S_m$, where $S_m$ is the $m^{\rm th}$ pixel's domain.      

\subsection{Sparse Born iterative method} \label{sec:sbim}
In \cite{desmal2014shrinkage}, a sparse approach of BIM (SBIM) was proposed based on ISTA. SBIM is summarized in the following algorithm. 

\begin{algorithm}[H]
\caption{Sparse Born Iterative Method}
\begin{algorithmic}[1]
\renewcommand{\algorithmicrequire}{\textbf{Input:}}
\REQUIRE $l_{1}$ 
\STATE \textit{\rm{Initialization:} } $\{ {\rm \bf e}^{\rm tot}_{tr,1}={\rm \bf e}^{\rm inc}_{tr}, \; \forall{tr} \}$
\FOR {$i = 1$ to $N^{\rm bim}$}
\STATE ${\rm \bf H}_{i}=cas({\rm \bf G}{\mathcal D}({\rm \bf e}^{\rm tot}_{1,i}),...,{\rm \bf G}{\mathcal D}({\rm \bf e}^{\rm tot}_{N^{tra},i}))  $
\STATE ${\rm \bf t}_{i}=min_{\rm \bf t}  || {\rm \bf H}_{i} {\rm \bf t}- {\rm \bf e}^{\rm mea}  ||^2_2$\; s.t. $| {\rm \bf t} | < l_1$
\FOR {$tr = 1$ to $N^{\rm tra}$}
\STATE ${\rm \bf e}^{\rm tot}_{tr,i+1}=\left ({\rm \bf I}+ {\rm \bf A} {\mathcal D}({\rm \bf t}_{i}) \right )^{-1} {\rm \bf e}^{\rm inc}_{tr} $
\ENDFOR
\ENDFOR
\RETURN ${\bar{\bf{t}}}_{N^{\rm bim}}$
\end{algorithmic} \label{alg:NLWKZ}
\end{algorithm}
Step~1 performs Born approximation by setting the total fields $ {\rm \bf e}^{\rm tot}_{tr,1}$ equal to the incident fields ${\rm \bf e}^{\rm inc}_{tr}$ for all the transmitters. Step~2 starts the SBIM iteration loop that iterates up to $N^{\rm bim}$ counts. The function $cas(.)$ in Step~3 cascades ${\rm \bf G}{\mathcal D}({\rm \bf e}^{\rm tot}_{tr,i})$ for all the transmitters to end up with a single linear ill-posed operator ${\rm \bf H}_{i}$ known as the observation matrix. Step~4 contains a linear sparse optimization problem that minimizes a least-square term of the data misfit, which is constraint by a first-norm penalty, where $l_1$ is the first-norm level of the contrast vector ${\rm \bf t}$. The measurement field vector ${\rm \bf e}^{\rm mea}={\rm \bf e}^{\rm sca}+\eta$, where ${\rm \bf e}^{\rm sca}$ cascades the scatter fields for all the transmitters, while $\eta$ is an additive white Gaussian noise. The optimization problem in Step~4 is carried out using the ISTA algorithm. ISTA iteration reads           
\begin{equation} \label{eq:lwb}
{\rm \bf t}_{i,l}={\mathcal T}^{\delta}\left ( {\rm \bf t}_{i,l-1}-\gamma_{i} {\rm \bf H}_{i}^{\dagger} \left ( {\rm \bf H}_{i}{\rm \bf t}_{i,l-1}-{\rm \bf e}^{\rm mea}   \right ) \right )
\end{equation}
where ${\mathcal T}^{\delta}(.)$ is the complex soft-thresholding and $\gamma_{i}$ is the ISTA step length \cite{daubechies2004iterative, desmal2014shrinkage}. ISTA iteration index ``$l$" runs up to $N^{\rm lwb}$ counts. Step~5 to 6 implements a loop that updates the total fields for all the transmitters using (\ref{eq:stat}) and the updated ${\rm \bf t}_{i}$ in Step~4.    

\subsection{Trained Born iterative method} \label{sec:tbim} 
\begin{figure}[!t]
\centering
\includegraphics[width=0.99\columnwidth]{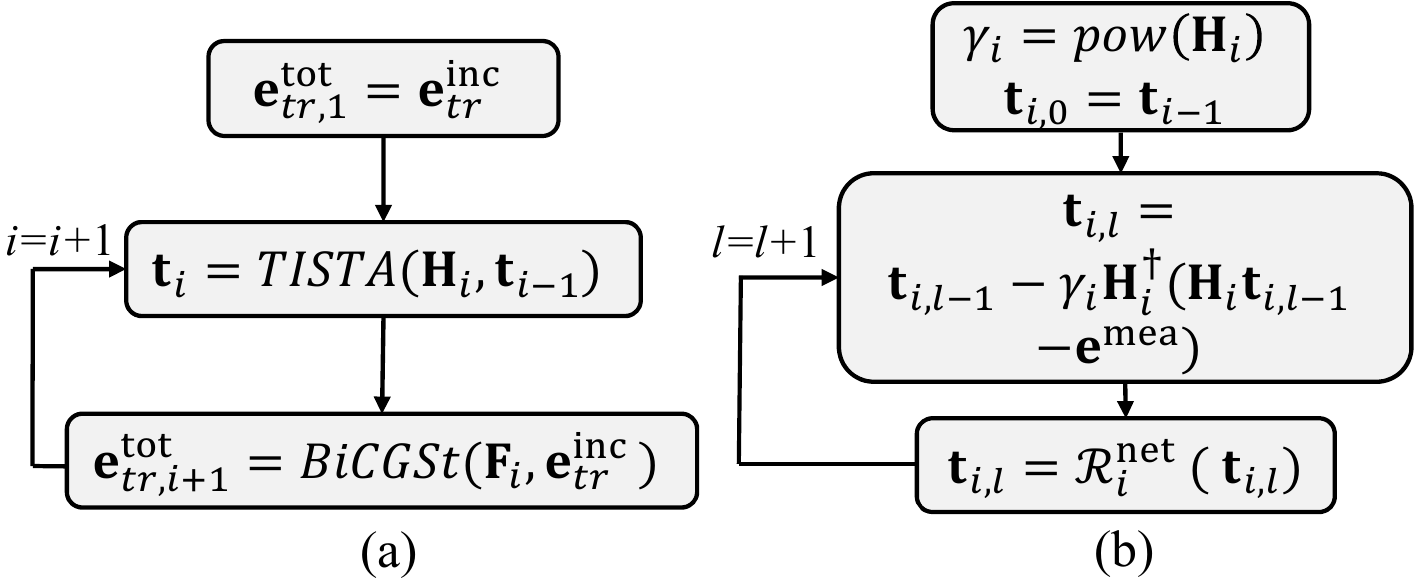}
\caption{(a) TBIM model, and (b) TISTA block.}
\label{fig:tbim}
\end{figure}
Figure~\ref{fig:tbim}(a) shows the TBIM. The block on the top applies the first-order Born approximation (BA). Then a loop executes the trained version of ISTA (TISTA) and updates ${\rm \bf e}_{tr,i+1}^{\rm tot}$ consequently. The inversion in Step~4 of the SBIM algorithm is carried out using the biconjugate gradient stabilized (BiCGSt) algorithm that runs over fixed $N^{\rm bcg}$ iteration counts, where the matrix ${\rm \bf F}_{i}={\rm \bf I}+ {\rm \bf A} {\mathcal D}({\rm \bf t}_{i})$. Fig.~\ref{fig:tbim}(b) shows the TISTA block in more detail. First, the power iteration is executed using fixed $N^{\rm pow}$ counts that compute $\gamma_i=1/\sigma_i^2$, where $\sigma_i$ is the maximum singular value of ${\rm \bf H}_i$. The latter approach normalizes TISTA step updates. Due to the BA assumption in the first TBIM step, the observation matrix ${\rm \bf H}_i$ becomes example independent. Meanwhile, during the second TBIM iteration step and after, ${\rm \bf H}_i$ becomes example-dependent as it gets updates corresponding to the examples' total fields, e.g. ${\rm \bf e}_{tr,i}^{\rm tot}$ for $i \ge 2$. In TISTA frameworks, the step length $\gamma_i$ is considered a training parameter, while the linear operator is fixed over the training set \cite{jin2017deep,ongie2020deep,gilton2019neumann}. Therefore, training TISTA during the first TBIM step considering $\gamma_i$ as a training parameter works as the assumptions above hold. On the other hand, the TISTA training process diverges throughout the second TBIM iteration step and after because these assumptions are not satisfied. The previous argument holds in Newton and quasi-Newton frameworks as well. In this work, the power iteration normalizes TISTA iteration updates, which makes TISTA training applicable.
\begin{figure}[!t]
\centering
\includegraphics[width=0.99\columnwidth]{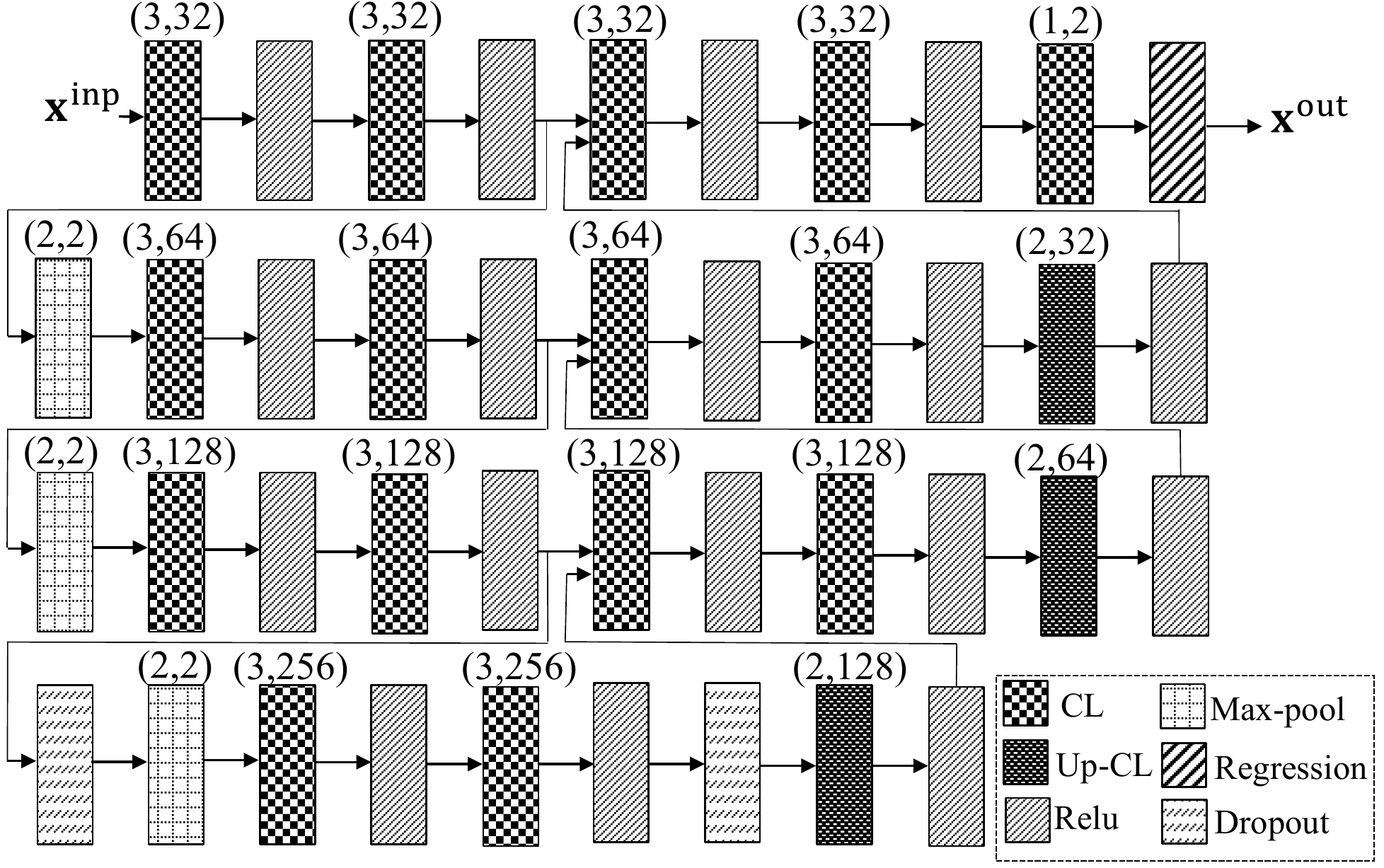}
\caption{U-net based regularization network ${\mathcal R}_{i}^{\rm net}({\rm \bf t})$. The numbers over convolutional layers (CLs) indicate the window size and number of filters, respectively. The numbers over max-pooling (max-pool) layers indicate the down-sampling window sizes. The numbers over up-convolutional layers (Up-CL) indicate the up-sampling window size and the number of filters, respectively.}
\label{fig:regNet}
\end{figure}

The network ${\mathcal R}_{i}^{\rm net}(.)$ in Fig.~\ref{fig:tbim}(b) was designed based on the well-known U-net architecture, as shown in Fig.~(\ref{fig:regNet}). The network input ${\rm \bf x}^{\rm inp}$ contains two channels of real and imaginary parts of contrast ${\rm \bf t}_{i,l}$. The output ${\rm \bf x}^{\rm out}$ is the denoised/filtered image of ${\rm \bf x}^{\rm inp}$. The network ${\mathcal R}_{i}^{\rm net}(.)$ was trained per TBIM iteration ``$i$", while mean square error of the reconstructed contrast profiles was implemented as a cost function. In the first TBIM iteration, $i=1$, ${\mathcal R}_{i}^{\rm net}(.)$ weights were randomly initialized. Then the subsequent ${\mathcal R}_{i}^{\rm net}(.)$ networks were initialized using the trained weights of ${\mathcal R}_{i}^{\rm net}(.)$ from the previous TBIM iteration step. The numerical results section illustrates high-quality image restoration after running a few TBIM iteration steps, e.g., three steps. The suggested network training procedure minimizes the memory allocation, which is a bottleneck in TISTAs.                    

\section{Numerical results} \label{sec:results}
\begin{figure}[!t]
\centering
\includegraphics[width=0.7\columnwidth]{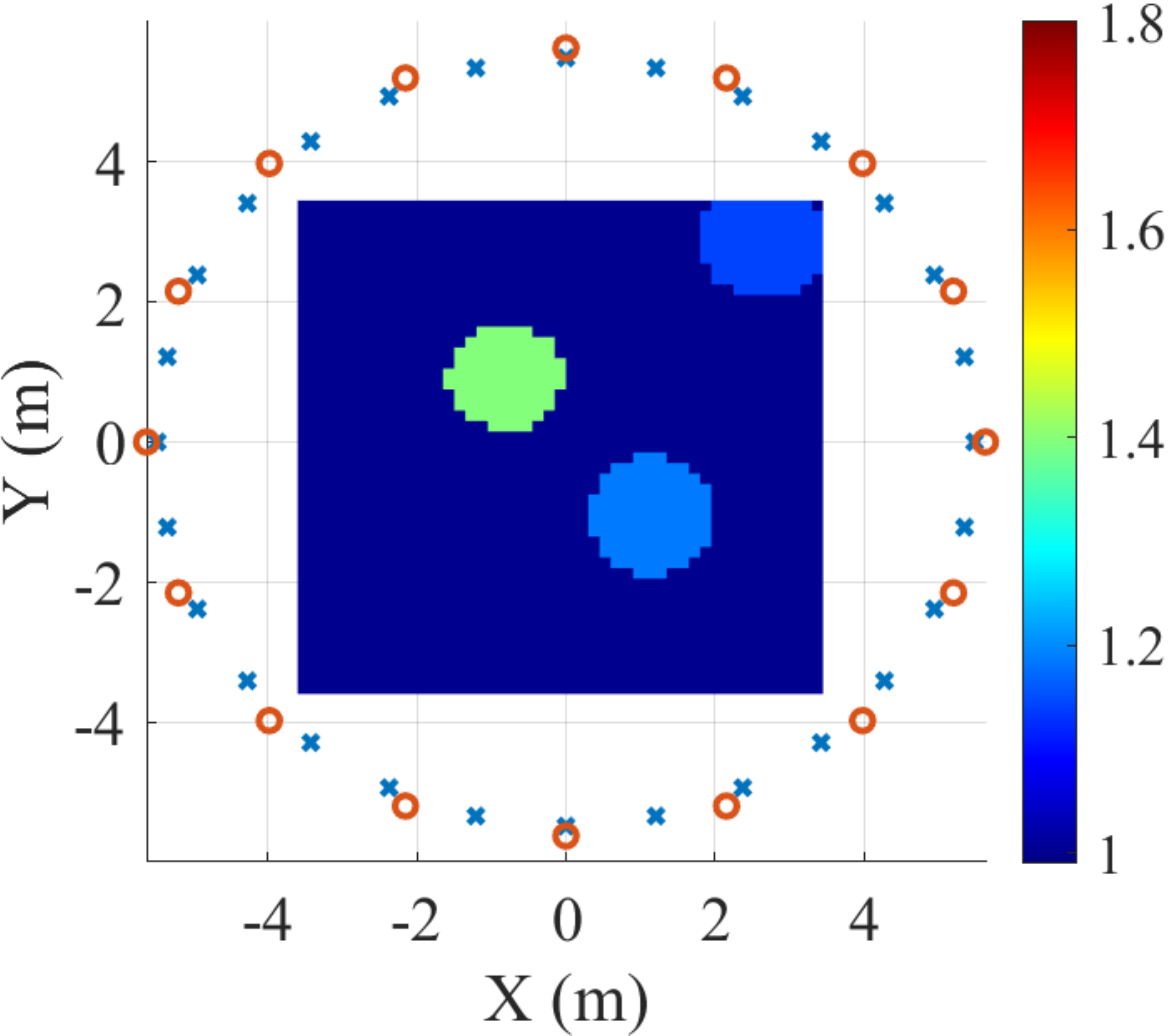}
\caption{Problem configuration.}
\label{fig:config}
\end{figure}
A set of 70000 examples was created that contains cylindrical target(s). 64000 of them were used for training, 2000 for validation, and 2000 for testing. The number of targets was randomized that can reach up to three targets per example. The target's radius, center location, contrast level (ranged from 0.1 to 0.9) were randomized. Fig.~\ref{fig:config} shows the actual contrast profile of the first example in the set along with the receiver-transmitter locations. The colorbar scale shown in Fig.~\ref{fig:config} will be used on the following examples. The operating frequency is 110~MHz while the pixel's side length $\Delta d=0.15$~m. The relative norm error (RNE) is implemented as a criterion to measure the performance of examples. Let ${\rm \bf x}^{\rm a}$ be a reference vector while ${\rm \bf x}$ be the vector under test. The RNE is defined as follows
\begin{equation} \label{eq:RNE}
{\rm RNE}=\frac{|| {\rm \bf x} - {\rm \bf x}^{\rm a} ||^2_2}{|| {\rm \bf x}^{\rm a} ||^2_2}\times 100 \%
\end{equation}
In case ${\rm \bf x}^{\rm a}$ and ${\rm \bf x}$ consist of multiple exmaples (set based), the mean RNE (MRNE) is used as a criteria. The examples and networks training in this work were coded using Pytorch v1.8.1 and executed using GTX 1080 GPU, Intel i7 CPU, and 16 Gigabyte RAM.   

\begin{figure*}[!t]
\centering
\subfloat{\includegraphics[width=0.28\columnwidth]{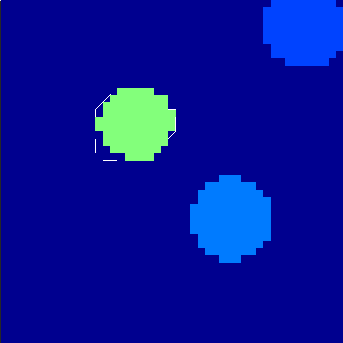}}
\hfil
\subfloat{\includegraphics[width=0.28\columnwidth]{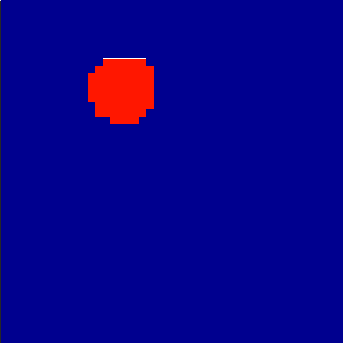}}
\hfil
\subfloat{\includegraphics[width=0.28\columnwidth]{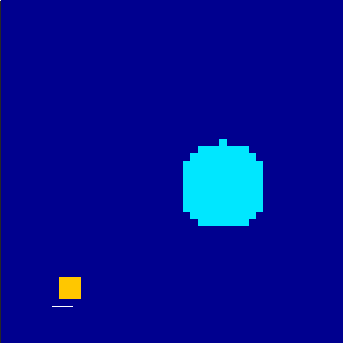}}
\hfil
\subfloat{\includegraphics[width=0.28\columnwidth]{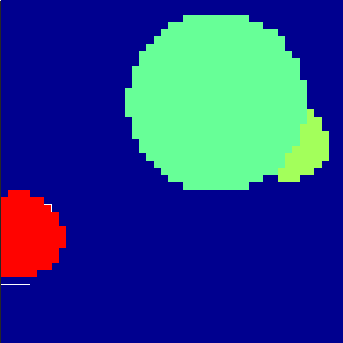}}
\hfil
\subfloat{\includegraphics[width=0.28\columnwidth]{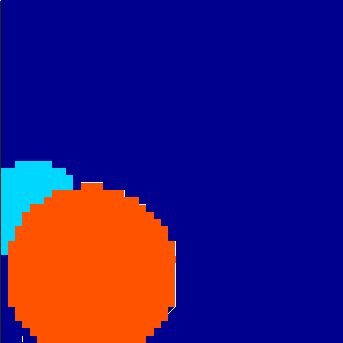}}
\hfil
\subfloat{\includegraphics[width=0.28\columnwidth]{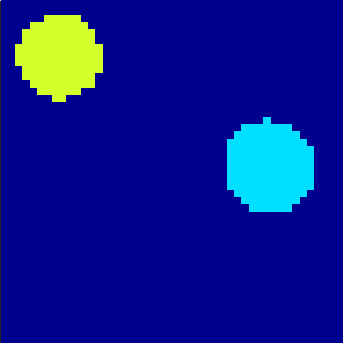}}
\hfil
\subfloat{\includegraphics[width=0.28\columnwidth]{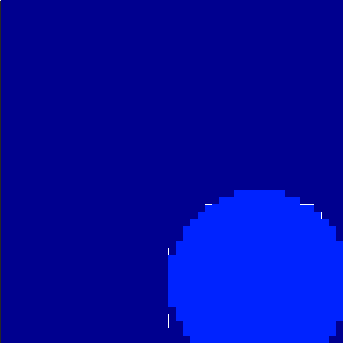}}
\\ 
\vspace{-0.25cm}
\subfloat{\includegraphics[width=0.28\columnwidth]{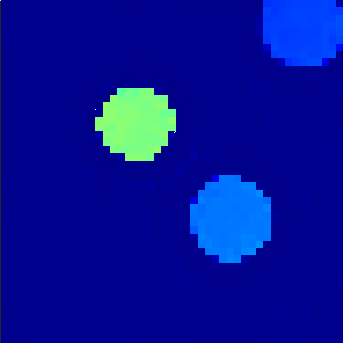}}
\hfil
\subfloat{\includegraphics[width=0.28\columnwidth]{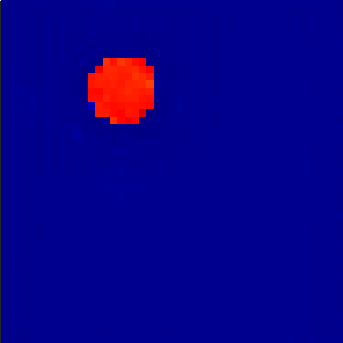}}
\hfil
\subfloat{\includegraphics[width=0.28\columnwidth]{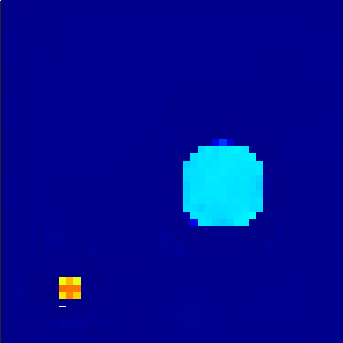}}
\hfil
\subfloat{\includegraphics[width=0.28\columnwidth]{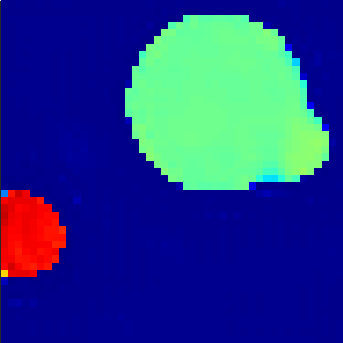}}
\hfil
\subfloat{\includegraphics[width=0.28\columnwidth]{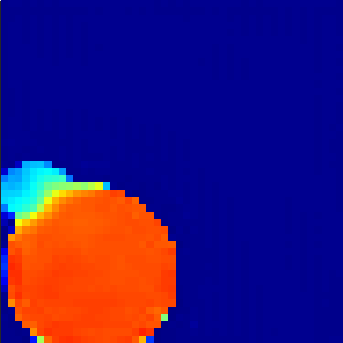}}
\hfil
\subfloat{\includegraphics[width=0.28\columnwidth]{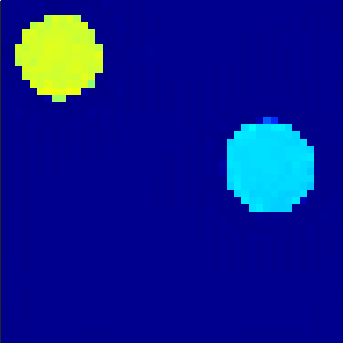}}
\hfil
\subfloat{\includegraphics[width=0.28\columnwidth]{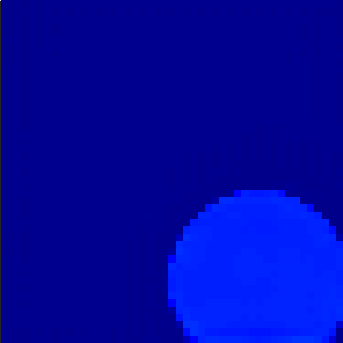}} 
\\
\vspace{-0.25cm}
\subfloat{\includegraphics[width=0.28\columnwidth]{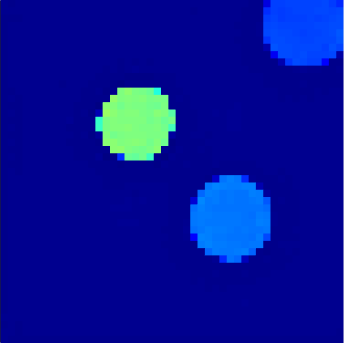}}
\hfil
\subfloat{\includegraphics[width=0.28\columnwidth]{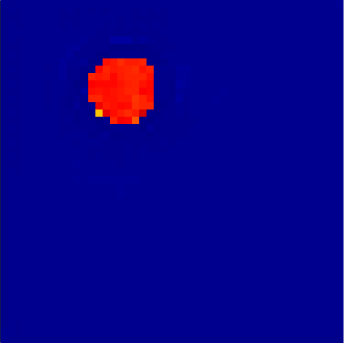}}
\hfil
\subfloat{\includegraphics[width=0.28\columnwidth]{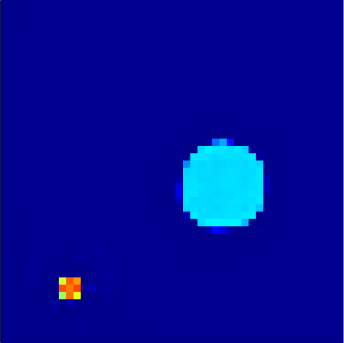}}
\hfil
\subfloat{\includegraphics[width=0.28\columnwidth]{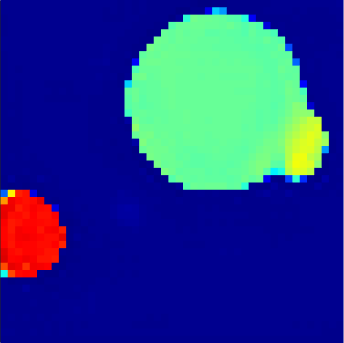}}
\hfil
\subfloat{\includegraphics[width=0.28\columnwidth]{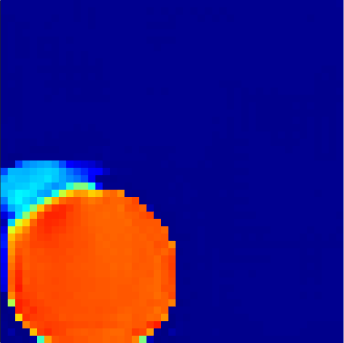}}
\hfil
\subfloat{\includegraphics[width=0.28\columnwidth]{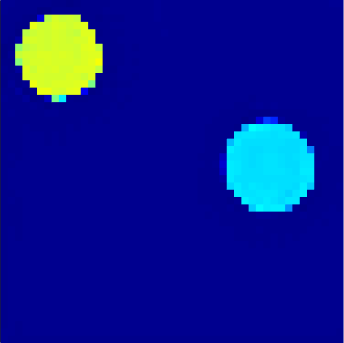}}
\hfil
\subfloat{\includegraphics[width=0.28\columnwidth]{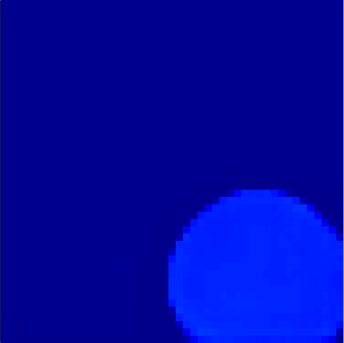}} 
\\
\vspace{-0.25cm}
\subfloat{\includegraphics[width=0.28\columnwidth]{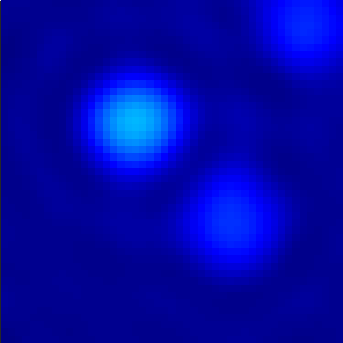}}
\hfil
\subfloat{\includegraphics[width=0.28\columnwidth]{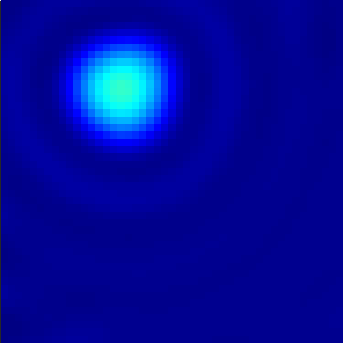}}
\hfil
\subfloat{\includegraphics[width=0.28\columnwidth]{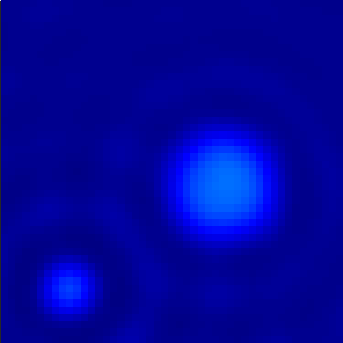}}
\hfil
\subfloat{\includegraphics[width=0.28\columnwidth]{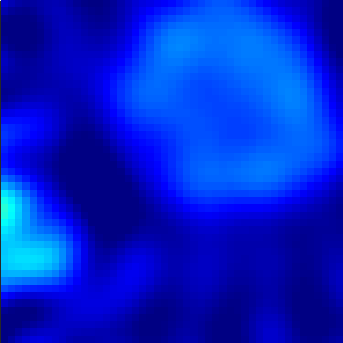}}
\hfil
\subfloat{\includegraphics[width=0.28\columnwidth]{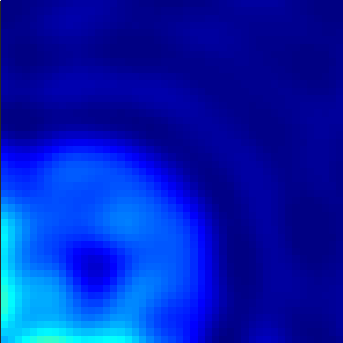}}
\hfil
\subfloat{\includegraphics[width=0.28\columnwidth]{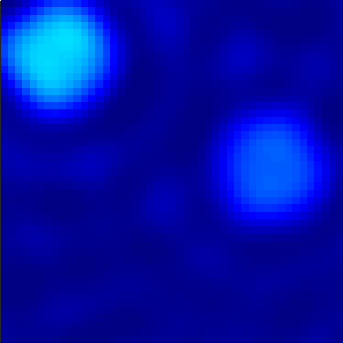}}
\hfil
\subfloat{\includegraphics[width=0.28\columnwidth]{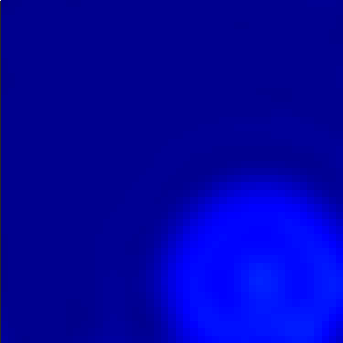}} 
\caption{The first row shows the actual profiles of the first seven examples in the testing set. The second, third, and fourth rows show the noiseless TBIM, 25 dB TBIM, and noiseless SBIM reconstructed profiles, respectively.  }
\label{fig:exmaples}
\end{figure*}

\begin{table}[]
\centering
\caption{RNEs of the examples shown in Fig.~\ref{fig:exmaples}.}
\label{tab:examples}
\begin{tabular}{|c|c|c|c|c|c|}
\hline
 Ex. No.     & \begin{tabular}[c]{@{}c@{}}Noiseless\\ TBIM\end{tabular} & \begin{tabular}[c]{@{}l@{}}25 dB\\ TBIM\end{tabular} & \begin{tabular}[c]{@{}c@{}}15 dB\\ TBIM\end{tabular} & \begin{tabular}[c]{@{}c@{}}10 dB\\ TBIM\end{tabular} & \begin{tabular}[c]{@{}c@{}}Noiseless\\ SBIM\end{tabular} \\ \hline
Ex. 1 & 6.55\%   & 11.40\%    & 15.64\%     & 17.13\%   & 62.27\%   \\ \hline
Ex. 2 & 2.34\%   & 3.56\%     & 13.07\%     & 18.71\%   & 70.71\%   \\ \hline
Ex. 3 & 7.96\%   & 11.85\%    & 12.68\%     & 16.07\%   & 64.53\%   \\ \hline
Ex. 4 & 5.73\%   & 10.44\%    & 13.69\%     & 21.14\%   & 82.16\%   \\ \hline
Ex. 5 & 10.29\%  & 10.70\%    & 15.08\%     & 15.66\%   & 95.08\%   \\ \hline
Ex. 6 & 4.05\%   & 6.40\%     & 12.43\%     & 14.98\%   & 64.11\%   \\ \hline
Ex. 7 & 6.65\%   & 9.75\%     & 14.12\%     & 11.28\%   & 35.40\%   \\ \hline
\end{tabular}
\end{table}

\begin{table}[]
\centering
\caption{Testing set MRNE per TBIM iteration count at different SNR levels.}
\label{tab:MRNE}
\begin{tabular}{|c|c|c|c|}
\hline
   SNR      & $i=1$     & $i=2$     & $i=3$     \\ \hline
Noiseless & 13.74\% & 9.46\%  & 7.64\%  \\ \hline
25 dB     & 14.54\% & 11.56\% & 11.17\% \\ \hline
15 dB     & 16.70\% & 14.55\% & 14.13\% \\ \hline
10 dB     & 19.11\% & 17.31\% & 17.26\% \\ \hline
\end{tabular}
\end{table}
Figure~\ref{fig:exmaples} shows the reconstructed profiles of noiseless TBIM, 25~dB TBIM, and noiseless SBIM for the first seven examples in the testing set, where $N^{\rm bim}=3$, $N^{\rm lwb}=6$, $N^{\rm bcg}=4$, $N^{\rm pow}=5$, and for the SBIM $\delta=0.001$. The RNEs of the reconstructed profiles in Fig.~\ref{fig:exmaples} are shown in Table~\ref{tab:examples}. Table~\ref{tab:MRNE} shows the MRNE of the testing set under different SNRs and TBIM iteration counts. The reconstructed profiles from the TBIM at different SNRs outperform the conventional noiseless SBIM.

\ifCLASSOPTIONcaptionsoff
  \newpage
\fi



%



\bibliographystyle{IEEEtran}
\bibliography{References}

%








\end{document}